\documentclass[prl,aps,preprint,amssymb]{revtex4}

\usepackage{epsfig}

\def\be{\begin{equation}}
\def\ee{\end{equation}}
\def\mueg{\mu^- \to e^- \gamma}
\def\taueg{\tau^- \to e^- \gamma}
\def\taumug{\tau^- \to \mu^- \gamma}
\newcommand {\eq} [1] {Eq.~(\ref{#1})}
\newcommand {\eqz} [2] {Eqs.~(\ref{#1}) and (\ref{#2})}
\newcommand {\fig} [1] {Fig.~\ref{#1}}
\newcommand {\misset}  {E_T\hspace{-4mm}/ \hspace{1mm}}
\newcommand {\tab} [1] {Table~\ref{#1}}

\begin{document}

\title{Large lepton flavor violating signals in supersymmetric
       particle decays at future $e^+ e^-$ colliders.}
\author{W.~Porod$^{a,b}$ and W.~Majerotto$^{a}$\\[0.5cm] 
\small
$^a$ Inst.~f.~Hochenergiephysik, \"Oster.~Akademie d.~Wissenschaften,
      A-1050 Vienna, Austria \\ \small
$^b$ Inst.~f\"ur Theor. Physik, Universit\"at Z\"urich, CH-8057 Z\"urich,
      Switzerland
}

\begin{abstract}
  We study lepton flavor violating signals at a future $e^+ e^-$
  linear collider within the general MSSM, allowing for the most
  general flavor structure. We demonstrate that there is a large
  region in parameter space with large signals, while being
  consistent with present experimental bounds on rare lepton decays
  such as $\mueg$.  In our analysis, we include all
  possible signals from charged slepton and sneutrino
  production and their decays as well as from the decays of
  neutralinos and charginos. We also consider the background 
  from the Standard Model {\it and} the MSSM. We find that in general
  the signature $e \tau \misset$ is the most pronounced one. 
  We demonstrate that even for
  an integrated luminosity of 100 fb$^{-1}$ the signal can be large. 
  At a high luminosity linear collider, precision experiments will allow one
  to determine  the lepton flavor
  structure of the MSSM.
\end{abstract}
\pacs{Lxyz}
\maketitle

There are stringent constraints on lepton flavor violation (LFV) in the charged
lepton sector, the strongest coming from the decay branching ratio of
$\mueg$, $BR(\mueg) < 1.2 \times 10^{-11}$ \cite{Brooks:1999pu}. Others are
$BR(\mu^- \to e^- e^+ e^-) < 10^{-12}$, $BR(\taueg) < 2.7 \times 10^{-6}$,
$BR(\taumug) < 1.1 \times 10^{-6}$, see \cite{Groom:2000in}.

On the other hand, recent experiments \cite{Nu1} indicate that at
least $\nu_\mu$ and $\nu_\tau$ have an almost maximal mixing angle,
$\sin^2 2 \theta_{atm} > 0.88$. The latest results of SNO \cite{Nu2}
suggest that also the $\nu_e-\nu_\mu$ sector contains a large
mixing, whereas the third mixing angle has to be small \cite{Nu3}.
The Standard Model can account for the lepton flavor 
conservation in the charged lepton sector, but has to be extended to
account for neutrino masses and mixings, e.g.~by the see-saw mechanism
and by introducing heavy right-handed Majorana neutrinos
\cite{seesaw}.

In general, a gauge and supersymmetric invariant theory does neither conserve
total lepton number $L=L_e + L_\mu + L_\tau$ nor individual lepton number
$L_e$,  $L_\mu$ or  $L_\tau$. One usually invokes R-parity symmetry, which
forces total lepton number conservation but still allows the violation of
individual lepton number, e.g.~due to loop effects in $\mueg$ \cite{ref11}.
The Minimal Supersymmetric Standard Model (MSSM) with R-parity
conservation embedded in a GUT theory induces LFV
\cite{ref6,ref6a,ref9} at the weak scale. This is a consequence of having
leptons and quarks in the same GUT multiplet and of the quark flavor
mixing due to the CKM matrix.  A general analysis of flavor changing
neutral (FCNC) effects in K- and B-meson as well as in lepton physics
was recently performed in \cite{ref7}.

Moreover, in the MSSM a large $\nu_\mu$-$\nu_\tau$ mixing can lead to a 
large  $\tilde \nu_\mu$-$\tilde \nu_\tau$ mixing
via renormalisation group equations \cite{ref10}. This leads to clear LFV
signals in slepton and sneutrino production
and in the decays of neutralinos and charginos into sleptons and sneutrinos
at the LHC \cite{ref12} and in $e^+ e^-$ and $\mu^+ \mu^-$ collisions 
\cite{ref8,ref13}. Signatures due to $\tilde e_R$-$\tilde \mu_R$ mixing
were discussed in \cite{ref14}. In all these studies, it has been assumed that
only one lepton flavor violating term dominates.

In this letter, we study the consequences of LFV
in the sfermion sector at future $e^+ e^-$ colliders, respecting present 
bounds on rare lepton decays. Assuming the {\it most general} mass matrices for
sleptons and sneutrinos, we demonstrate that
large signals are expected.

The most general charged slepton mass matrix including left-right mixing
as well as flavor mixing is given by:
\begin{equation}
  M^2_{\tilde l} = \left(
    \begin{array}{cc}
      M^2_{LL} &  M^{2\dagger}_{LR} \\
      M^2_{LR} &  M^2_{RR} \\
     \end{array} \right) \, ,
  \label{eq:sleptonmass}
\end{equation}
where the entries are $3 \times 3$ matrices. They are given by
\begin{eqnarray}
  \label{eq:massLL}
  M^2_{LL,ij} &=& M^2_{L,ij} + \frac{ v^2_d Y^{E*}_{ki} Y^{E}_{kj} }{2}
  \nonumber \\ &&
  + \frac{\left( {g'}^2 -  g^2 \right) (v^2_d - v^2_u) \delta_{ij}}{8}  \, ,\\
  \label{eq:sleptonmassLR}
  M^2_{LR,ij} &=& \frac{ v_d A^*_{ji} - \mu v_u Y^E_{ij} }{\sqrt{2}}  \, ,\\ 
  M^2_{RR,ij} &=& M^2_{E,ij} + \frac{ v^2_d Y^{E}_{ik} Y^{E*}_{jk} }{2}
      -  \frac{ {g'}^2  (v^2_d - v^2_u) \delta_{ij}}{4}  \, .%\\
\end{eqnarray}
The indices $i,j,k=1,2,3$ characterize the flavors $e,\mu,\tau$.
$M^2_{LL}$ and $M^2_{RR}$ are the soft SUSY breaking mass matrices for
left and right sleptons, respectively. $A_{ij}$ are the trilinear soft
SUSY breaking couplings of the sleptons and Higgs boson.
The physical mass eigenstates states $\tilde l_n$ are given by 
$\tilde l_n = R^{\tilde l}_{nm} \tilde l'_m$ with 
$l'_m = (\tilde e_L, \tilde \mu_L, \tilde \tau_L,
          \tilde e_R, \tilde \mu_R, \tilde \tau_R)$.
Similarly, one finds for the sneutrinos
\begin{eqnarray}
  M^2_{\tilde \nu,ij} &=&  M^2_{L,ij} 
  + \frac{ \left( g^2 + {g'}^2 \right) (v^2_d - v^2_u) \delta_{ij}}{8}
  \label{eq:sneutrinomass}
\end{eqnarray}
with the physical mass eigenstates 
$\tilde \nu_i = R^{\tilde \nu}_{ij}\tilde \nu_j'$ and 
$\tilde \nu_j' = (\tilde \nu_e, \tilde \nu_\mu, \tilde \nu_\tau) $.
The relevant interactions for this study are given by:
\begin{eqnarray}
  \label{eq:CoupChiSfermion}
 {\cal L} &=& \bar l_i ( c^L_{ikm} P_L + c^R_{ikm} P_R)
               \tilde \chi^0_k \tilde l_m  
\nonumber \\
    &+&  \bar{l_i} (d^L_{ilr} P_L + d^R_{ijr} P_R)
           \tilde \chi^-_l \tilde{\nu}_r
\nonumber \\
 &+& \bar{\nu_i} (e^L_{ilm} P_L + e^R_{ilm} P_R) \tilde \chi^+_l \tilde{l}_m 
\, .
\end{eqnarray}
The specific form of the couplings $c^L_{ikm}$, $c^R_{ikm}$,
$d^L_{ikm}$, $d^R_{ikm}$, $e^L_{ikm}$ and $e^R_{ikm}$ will be given
elsewhere \cite{project2}. The first two terms in \eq{eq:CoupChiSfermion}
give rise to the
signals whereas the last one will give rise to the SUSY background.
%with
%\begin{eqnarray}
%%
% c^L_{ikm} &=&
%  - \sqrt{2} g' ( R^{\tilde l}_{m,i+3} )^*  N_{k1}^*
%  - ( R^{\tilde l}_{mi} )^*  Y^E_{ii} N_{k3}^* \\
% c^R_{ikm} &=&
%  ( R^{\tilde l}_{ki} )^* \frac{g'  N_{k1} + g N_{k2} }{\sqrt{2}}
%   - ( R^{\tilde l}_{m,i+3} )^*  Y^E_{ii} N_{j3} \\ 
%%
% d^L_{ilr} &=& Y^E_{ii} ( R^{\tilde \nu}_{ri} )^* U^*_{l2} \\
% d^R_{ilr} &=& - g * ( R^{\tilde \nu}_{ri} )^*  V_{l1} \\ 
%%
% e^L_{ilm} &=& 0 \\
% e^R_{ilm} &=& \sum_r \left( - g ( R^{\tilde l}_{mr} )^* R^\nu_{ir} U_{l1}
%         + Y^E_{rr} ( R^{\tilde l}_{m,r+3} )^* R^\nu_{ir} U^*_{l2} \right)
%\end{eqnarray}
%where we have chosen the basis where the charged lepton Yukawa coupling
%is diagonal $Y^E_{ij} = \sqrt{2} m_i / v_d \delta_{ij}$. $R^\nu_{ij}$ is
%the neutrino mixing matrix, $N$ diagonalises the neutralino mass matrix 
%in the basis  $\tilde B, \tilde W_3, \tilde H^0_u, \tilde H^0_d$ and
%$U$ and $V$ are the mixing matrices of the charginos.

As mentioned above, most studies so far consider the case where only one
of the flavor mixing entries in the slepton (\eq{eq:sleptonmass}) and 
sneutrino mass (\eq{eq:sneutrinomass}) matrices is non-zero, as for
instance, $M^2_{L,23} \ne 0$. It is the purpose of this study to allow
for all possible flavor violating entries  in 
\eqz{eq:sleptonmass}{eq:sneutrinomass} which are compatible
with the present bounds on lepton number violating processes, such as
$\mueg, e^- e^+ e^-$, $\taueg$, $\taumug$ and 
$Z\to e \mu, e \tau, \mu \tau$.
For definiteness, we have taken the first of the mSUGRA points of Snowmass' 01 
\cite{Georg} as
reference point which is characterized by $M_{1/2} = 250$~GeV,
$M_0=100$~GeV, $A'_0=-100$~GeV, $\tan \beta = 10$ and sign$(\mu)=+$.
Note that $A'_0$ has  to be multiplied by the Yukawa couplings to get
the $A$ parameter as given in \eq{eq:sleptonmassLR}.
The corresponding parameters at the scale 
$Q=\sqrt{M_{Q3} M_{U3}}$ are given in \tab{tab:par} and the physical
masses (computed at one-loop) in \tab{tab:mass}. We keep all parameters fixed
except for the slepton parameters $M^2_L$, $M^2_R$ and $A_l$ where all
entries are varied in the whole range compatible with 
the experimental constraints.

\begin{table}
\caption{SUSY parameters at the scale $Q=\sqrt{M_{Q3} M_{U3}}$ for
$M_{1/2} = 250$~GeV,
$M_0=100$~GeV, $A'_0=-100$~GeV, $\tan \beta = 10$ and sign$(\mu)=+$.}
\label{tab:par}
\begin{center}
\begin{tabular}{rrr}
 $M_1$ = 107.9 & $M_2$ =   208.4 & $M_3$ =  611.6    \\ \hline
 $M_{E_1}$ = 138.7 & $M_{L_1}$ = 202.3 &  $A_e/Y_e$  = -257.3  \\ 
 $M_{E_3}$ = 136.3 & $M_{L_3}$ = 201.5 & $A_\tau/Y_\tau$  = -257.3    \\ \hline
 $M_{D_1}$ =  536 & $M_{U_1}$ =  540 & $M_{Q_1}$ =  562     \\
 $M_{D_3}$ =  534 & $M_{U_3}$ =  427 & $M_{Q_3}$ =  509 \\
 $A_{b}$ = -863 &  $A_{t}$ = -503 
\end{tabular}
\end{center}
\end{table}

\begin{table}
\caption{SUSY spectrum  for
$M_{1/2} = 250$~GeV,
$M_0=100$~GeV, $A_0=-100$~GeV, $\tan \beta = 10$ and sign$(\mu)=+$.}
\label{tab:mass}
\begin{center}
\begin{tabular}{cccc}
 $m_{h^0}$ = 111 & $m_{H^0} = 395$ & $m_{A^0} = 395$ & $m_{H^+}= 403$ \\ \hline
 $m_{\tilde g} = 618$ & &
 $m_{\tilde \chi^+_1} = 193.6$ & $m_{\tilde \chi^+_2} = 376.2$ \\ 
 $m_{\tilde \chi^0_1} = 103.1$ & $m_{\tilde \chi^0_2} = 194.6$  &
 $m_{\tilde \chi^0_3} = 355.1$ & $m_{\tilde \chi^0_4} = 376.0$ \\ \hline 
 $m_{\tilde e_R} = 146.9$ & $m_{\tilde e_L} = 214.7$ &
 $m_{\tilde \tau_1} = 138.6$ & $m_{\tilde \tau_2} = 217.7$ \\
 $m_{\tilde \nu_e} = 199.4$ & $m_{\tilde \nu_\tau} = 198.5$ & &  \\ \hline 
 $m_{\tilde d_R} = 560$ & $m_{\tilde d_L} = 585$ &   
 $m_{\tilde u_R} = 561$ & $m_{\tilde u_L} = 579$   \\
 $m_{\tilde b_1} = 530$ & $m_{\tilde b_2} = 559$ &   
 $m_{\tilde t_1} = 407$ & $m_{\tilde t_2} = 600$   
\end{tabular}
\end{center}
\end{table}

\begin{figure}
\setlength{\unitlength}{1mm}
\begin{picture}(70,98)
%\put(-3,-2){\mbox{\epsfig{figure=parameters.eps,height=10cm,width=8cm}}}
\put(-3,30){\mbox{\epsfig{figure=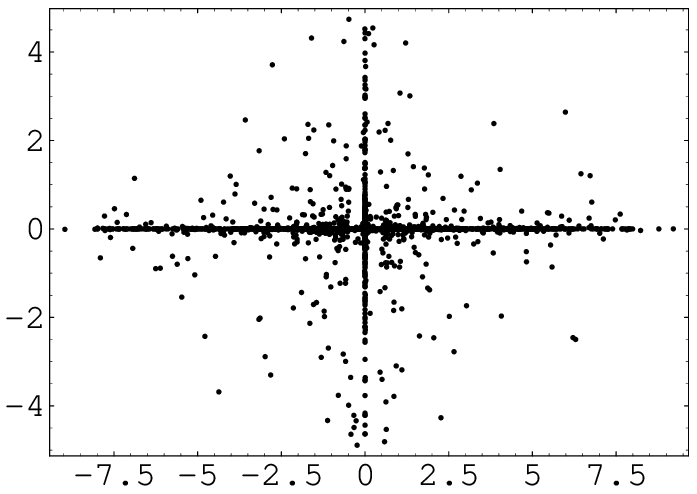,height=8.cm,width=8cm}}}
\put(-1,-21){\mbox{\epsfig{figure=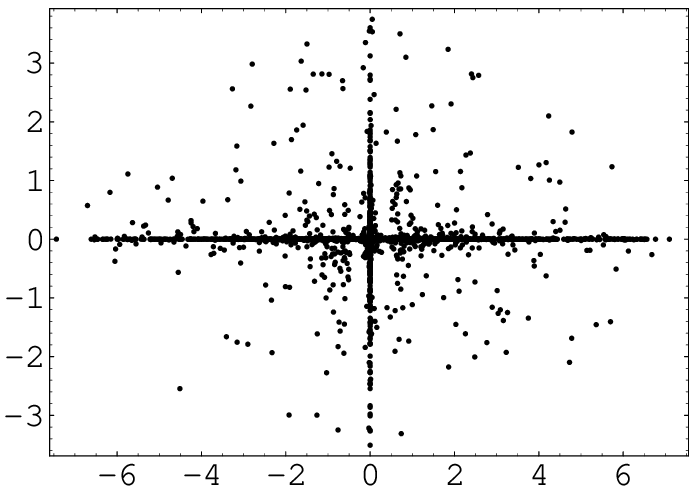,height=8.cm,width=8cm}}}
\put(-2,93){\mbox{\bf (a)}}
\put(4,92){\mbox{$M^2_{L,13} \cdot 10^3$~GeV$^2$}}
\put(50,46.5){\mbox{$M^2_{R,13} \cdot 10^3$~GeV$^2$}}
\put(-2,42.5){\mbox{\small \bf (b)}}
\put(4,41.5){\mbox{\small $M^2_{L,23} \cdot 10^3$~GeV$^2$}}
\put(50,-4){\mbox{\small $M^2_{R,23} \cdot 10^3$~ GeV$^2$}}
\end{picture}
\caption{Ranges for parameters inducing lepton number violation.}
\label{fig:parameter}
\end{figure} 

We find values for  $|M^2_{R,ij}|$ up to $8 \cdot 10^3$~GeV$^2$, $|M^2_{L,ij}|$
up to $6 \cdot 10^3$~GeV$^2$ and $|A_{ij} v_d|$ up to 650~GeV$^2$ compatible 
with the 
constraints. In most cases, one of the mass squared parameters is at least
one order of magnitude larger than all the others. However, there is a
sizable part in parameters where at least two of the off-diagonal parameters
have the same order of magnitude as shown in \fig{fig:parameter}.

In what follows, we concentrate on possible LFV signals at a 500~GeV
$e^+ e^-$ collider: $e \mu \,  \misset$, 
$e \tau \, \misset$, $\mu \tau \,  \misset$, as well as the 
possibility of two additional jets.  
 We consider the following SUSY processes: 
$e^+ e^- \to \tilde l^-_i \tilde l^+_j, \tilde \nu_{i'} \bar{\tilde \nu}_{j'},
\tilde \chi^0_k \tilde \chi^0_m, \tilde \chi^+_n \tilde \chi^-_o$ as
well as stop and Higgs production. We take into account all possible SUSY
 and Higgs cascade decays. We have taken into account ISR- and
SUSY-QCD corrections for the production cross sections.

The main sources for the LFV signal stem from production of sleptons,
sneutrinos and their decays, for example:
\begin{eqnarray}
e^+ e^- \to \tilde l^-_i \tilde l^+_j \to l^-_k l^+_m 2 \tilde \chi^0_1 \, .
\end{eqnarray}
 We have also included  the oscillation between flavors, being
important
in the case that $\Delta m^2 < m \Gamma$ \cite{ref8, Dine:2001cf}.

%For the signal we take into account slepton and
%sneutrino production and their decays as well neutralino decays into 
%leptons with different flavors: 
%$ e^+ e^- \to \tilde l^-_j \tilde l^+_k \to l^-_j l^+_k
%               \tilde \chi^0_r \tilde \chi^0_s$,
%$e^+ e^- \to \tilde \nu_j \bar{\tilde \nu}_k \to l^-_j l^+_k
%        \tilde \chi^+_r \tilde \chi^-_s$,
%$e^+ e^- \to \tilde \chi^0_t \tilde \chi^0_r 
%         \to  l^-_j l^+_k  \tilde \chi^0_t \tilde \chi^0_s$ with $j \ne k$.
For the background we take into account all possible SUSY cascade
decays faking the signal and the Standard Model background from
$W$-boson pair production, $t$-quark pair production and $\tau$-lepton
pair production. The SM background has been calculated with the program
Whizard \cite{Kilian:2001qz}.
A SUSY background reaction is, for example, the chain
$\tilde \chi^0_r \to l^-_j \nu_i \tilde \chi^+_s \to l^-_j \nu_i l^+_k
\nu_n \tilde \chi^0_m$. We have generated 8000
points consistent with the experimental, varying
the parameters randomly on a logarithmic scale: $ 10^{-8} \le |A_{ij}|
\le 50$~GeV, $ 10^{-8} \le M^2_{ij} \le 10^4$~GeV$^2$. 
About 1200 of these have at least one signal larger than 0.1~fb.  In
\tab{tab:signal} we present the maximal cross section for various
signals with a cross section larger then $10^{-2}$~fb. 
The cross section for $e \tau \misset$ can go up to 250 fb leading 
to about $10^5$ events with a luminosity of 500~fb$^{-1}$.
 In the case of
two leptons with different flavors and 2 jets, we have put a veto on
b-jets because of the large background stemming from $t$-quark
production.  One observes that the cross section for $\mu^\pm \tau^\mp
\misset$ is somewhat smaller than the cross section for $e^\pm \mu^\mp
\misset$ \, and $e^\pm \tau^\mp \misset$. The reason for this is that
$\tilde e \tilde e$ production is larger than 
$\tilde \mu \tilde \mu$ ($\tilde \tau \tilde \tau$)
 due to the additional t-channel contribution.

\begin{figure}
\setlength{\unitlength}{1mm}
\begin{picture}(70,118)
%\put(-2,-2){\mbox{\epsfig{figure=SignalTauE.eps,height=12cm,width=8.2cm}}}
\put(-3,59){\mbox{\epsfig{figure=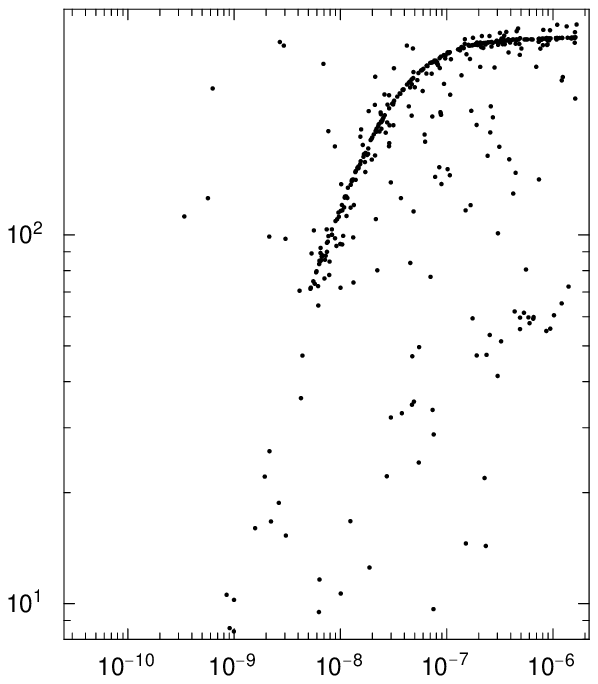,height=5.cm,width=7.5cm}}}
\put(-3,0){\mbox{\epsfig{figure=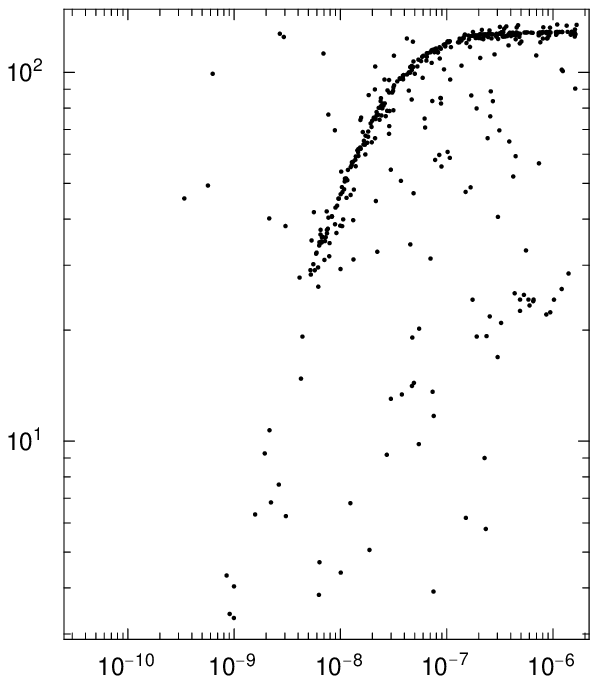,height=5.cm,width=7.5cm}}}
\put(-2,111.5){\mbox{\bf (a)}}
\put(5,111){\mbox{$\sigma$~[fb]}}
\put(58,57){\mbox{BR$(\tau \to e \gamma)$}}
\put(-2,52.5){\mbox{\bf (b)}}
\put(5,52){\mbox{$signal/\sqrt{background}$}}
\put(58,-2){\mbox{BR$(\tau \to e \gamma)$}}
\end{picture}
\caption{({\bf a}) Cross section in fb for the signal $e^\pm \tau^\mp$ + 
           missing transverse momentum  and  ({\bf b}) the ratio
           signal over square root of background  as a function of 
           BR$(\tau \to e \gamma)$ 
           for $\sqrt{s} = 500$~GeV, $P_{e^-} = 0$ and $P_{e^+} = 0$. In the
           latter case we have assumed an integrated luminosity of 
           100 fb$^{-1}$.}
\label{fig:signal1}
\end{figure}

In \fig{fig:signal1}a we show the cross section in fb of $e^+ e^- \to
e^\pm \tau^\mp \misset$ as a function of $BR(\taueg)$ and in
\fig{fig:signal1}b the ratio signal over square root of the background
($S/\sqrt{B}$) as a function $BR(\taueg)$ assuming an integrated
luminosity of 100~fb$^{-1}$.  Although no cuts have been applied,
there is in most cases a spectacular signal. The cases
where the ratio $S/\sqrt{B}$ is of order 1 or smaller should
clearly improve, once appropriate cuts are applied. For example, a cut
on the angular distribution of the final state leptons
will strongly reduce the $WW$ background. 
%
%A detailed study of possible cuts is
%beyond the scope of this letter and will presented elsewhere
%\cite{project2}. 
Further cuts as applied in the study of slepton
production \cite{vandervelde} will enhance the ratio  $S/\sqrt{B}$.

There is an accumulation of points in \fig{fig:signal1} along a band.
These points are characterized by large $\tilde e_R$-$\tilde \tau_R$
mixing which is less constraint by $\taueg$ than the corresponding
left-left or left-right mixing. 

% In the points above the line a
% cancellation between different contributions for $\taueg$ whereas in
% the points below the left-left and/or left-right mixing effects in
% $\taueg$ sum giving a somewhat smaller signal.

\begin{table}
\caption[]{Maximal cross section in fb for various signals at a 500 GeV 
$e^+ e^-$
collider. The $b$-jets have been excluded.}
\label{tab:signal}
\begin{center}
\begin{tabular}{ccccccc}
 & \multicolumn{5}{c}{(electron,positron polarization)} \\ 
signal & (0,0) & (-0.8,-0.6) & (-0.8,0.6) & (0.8,-0.6) & (0.8,0.6) \\ \hline
$e^\pm \mu^\mp \misset$ & 149 & 67 & 208 & 231 & 71 \\
$e^\pm \tau^\mp \misset$ & 178 & 93 & 220 & 248 & 72 \\
$\mu^\pm \tau^\mp \misset$ & 61 & 56 & 127 & 115 & 42 \\ \hline
$e^\pm \mu^\mp 2 j\misset$ & 0.13 & 0 & 0.01 & 0.38 & 0 \\
$e^\pm \tau^\mp 2 j\misset$ & 0.51 & 0 & 0.04 & 1.46 & 0 \\
$\mu^\pm \tau^\mp 2 j\misset$ & 0.5 & 0 & 0.04 & 1.43 & 0 \\ %\hline
%$e^\pm \mu^\mp 4 j\misset$ & 0.02 & 0 & 0  & 0.05 & 0 \\
%$e^\pm \tau^\mp 4 j\misset$ & 0.07 & 0 & 0  & 0.21 & 0 \\
%$\mu^\pm \tau^\mp 4 j\misset$ & 0.07 & 0 & 0 & 0.20 & 0 \\
\end{tabular}
\end{center}
\end{table}

\begin{figure}
\setlength{\unitlength}{1mm}
\begin{picture}(70,129)
%\put(-9,-2){\mbox{\epsfig{figure=SignalSqrtS.eps,height=12cm,width=8.5cm}}}
\put(-5,123){\mbox{\bf (a)}}
\put(-2,86){\mbox{\epsfig{figure=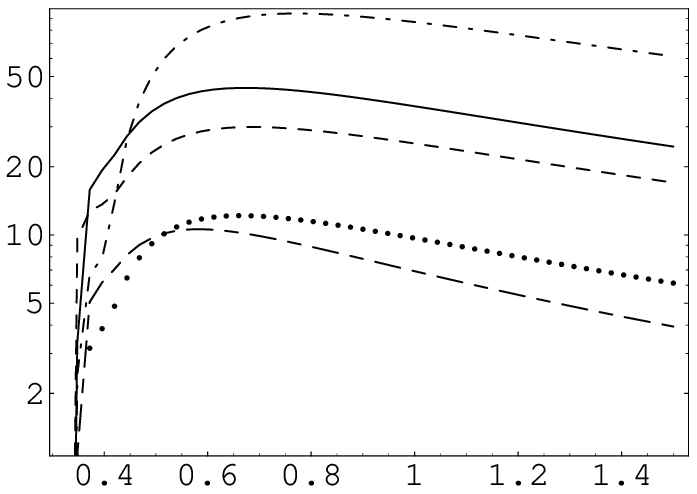,height=3.5cm,width=7.cm}}}
\put(1,122.5){\mbox{$\sigma(e^\pm \mu^\mp \misset)$~[fb]}}
\put(55,84){\mbox{$\sqrt{s}$~[TeV]}}
\put(-5,81.5){\mbox{\bf (b)}}
\put(-2,44){\mbox{\epsfig{figure=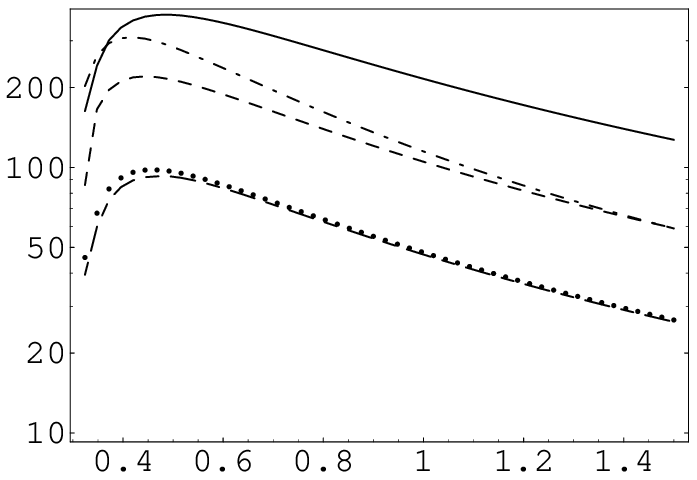,height=3.5cm,width=7.cm}}}
\put(1,81){\mbox{$\sigma(e^\pm \tau^\mp \misset)$~[fb]}}
\put(55,42){\mbox{$\sqrt{s}$~[TeV]}}
\put(-5,39.5){\mbox{\bf (c)}}
\put(-2,2){\mbox{\epsfig{figure=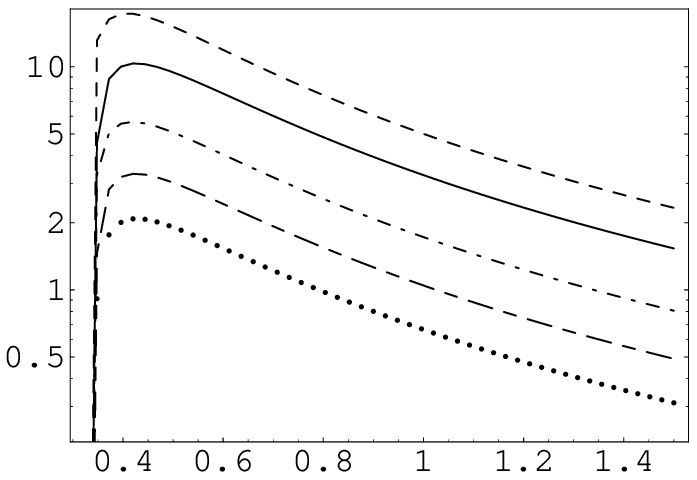,height=3.5cm,width=7.cm}}}
\put(1,39){\mbox{$\sigma(\mu^\pm \tau^\mp \misset)$~[fb]}}
\put(55,0){\mbox{$\sqrt{s}$~[TeV]}}
\end{picture}
\caption{The cross sections in fb for the signals
         $(e^\pm \mu^\mp \misset)$  ({\bf a}),
         $(e^\pm \tau^\mp \misset)$  ({\bf b}), and  
         $(\mu^\pm \tau^\mp \misset)$  ({\bf c}) as a function of $\sqrt{s}$
         for different beam polarizations. The different lines correspond to
         $(P_-, P_+) = (0,0)$ (full), 
         $(P_-, P_+) = (-0.8,-0.6)$ (dashed), 
         $(P_-, P_+) = (-0.8,0.6)$ (dashed dotted), 
         $(P_-, P_+) = (0.8,-0.6)$ (long dashed)  and
         $(P_-, P_+) = (0.8,0.6)$ (dotted). 
}
\label{fig:signal2}
\end{figure}

In \fig{fig:signal2} we study the dependence of the signal on the collider
energy for different beam polarizations. The various kinks are due to the onset
of the different production cross sections. One observes
a strong dependence on the beam polarization. Beam polarization is not
only useful for a possible reduction of the background, but might also
serve as a possible tool to disentangle different contributions to the signal.
We have chosen a point giving rise to a LFV signal in several
channels. The largest flavor violation entries are in this case
$M^2_{E,13} = 3440$~GeV$^2$, $M^2_{L,12} = -2.4$~GeV$^2$, $v_1 A_{31}
= 1.8$~GeV$^2$.  In this particular example, the branching ratios for
the rare lepton decays are within the reach of the next generation of
experiments, e.g.~BR($\taueg) = 2.6 \cdot 10^{-7}$.

For other points of the parameter space there will be only one or two
channels with large LFV signals.  However, the point chosen
demonstrates the general behavior of strong beam polarization dependence
of the various signals and is therefore quite representative. In the
case of additional jets in the final state, the cross section is lower
as can be seen from \tab{tab:signal}. At large values of $\sqrt{s}$
there are of course more open channels due to the production of
squarks. The corresponding LFV signals also show a pronounced
dependence on beam polarization \cite{project2}.

In conclusion, we have shown that the most general flavor violating structure
of the slepton and sneutrino mass matrix may lead to large lepton flavor
violating signals at a future $e^+ e^-$ collider
-- despite the strong constraints on rare lepton decays.

\acknowledgements

This work was supported by the `Fonds zur F\"orderung der
wissenschaftlichen Forschung' of Austria, project No.~P13139-PHY and
the Erwin Schr\"odinger fellowship Nr.~J2095, by the EU TMR Network
Contract No.~HPRN-CT-2000-00149, and partly by the Swiss 'Nationalfonds'.

\end{document}